%% file: main.tex
\documentclass[10pt,conference]{IEEEtran}
\IEEEoverridecommandlockouts

\usepackage{cite}
\usepackage{amsmath,amssymb,amsfonts}
\usepackage{algorithmic}
\usepackage{graphicx}
\usepackage{textcomp}
\usepackage{xcolor}
\usepackage{footnote}
\def\BibTeX{{\rm B\kern-.05em{\sc i\kern-.025em b}\kern-.08em
    T\kern-.1667em\lower.7ex\hbox{E}\kern-.125emX}}

\input{utils/macros}
\begin{document}
\bstctlcite{IEEEexample:BSTcontrol}

\title{ 
Toward Neurosymbolic Program Comprehension
}


\author{
\IEEEauthorblockN{
Alejandro Velasco, Aya Garryyeva, David N. Palacio, Antonio Mastropaolo, Denys Poshyvanyk}
\IEEEauthorblockA{
Department of Computer Science, William \& Mary \\
\{svelascodimate, lgarryyeva, danaderpalacio, amastropaolo, dposhyvanyk\}@wm.edu}
}


\maketitle

\input{sections/0_abstract}

\input{sections/1_introduction}
\input{sections/2_background}

\input{sections/2_approach}

\input{sections/3_case_study}

\input{sections/4_related}

\input{sections/5_future_plan}


\bibliographystyle{IEEEtran}
\bibliography{main}

\end{document}

%% file: utils/macros.tex
\usepackage{caption}
\usepackage{blindtext}
\usepackage{tcolorbox}
\usepackage[final]{pdfpages}
\usepackage{lipsum,multicol}
\usepackage{xcolor}
\usepackage{tikz}
\usepackage{listings}
\usepackage{enumitem}
\usepackage{hyperref}
\usepackage{amsfonts}
\usepackage{wrapfig}
\usepackage{subcaption} 
\usepackage{adjustbox}
\usepackage{colortbl}
\usepackage{fancybox}
\usepackage{multirow}
\usepackage[normalem]{ulem}
\useunder{\uline}{\ul}{}
\usepackage{enumitem}
\usepackage{float} 

\newcommand{\ie}{\textit{i.e.,}\xspace}
\newcommand{\eg}{\textit{e.g.,}\xspace}

\newcommand{\etal}{et al.\xspace}

\newtcolorbox{boxK}{
    fontupper = \small,
    sharpish corners, 
    boxrule = 0pt,
    toprule = 0pt, 
}

\newcommand{\secref}[1]{Sec.~\ref{#1}\xspace}
\newcommand{\figref}[1]{Fig.~\ref{#1}\xspace}
\newcommand{\tabref}[1]{Table~\ref{#1}\xspace}

\newcommand*\circled[1]{\tikz[baseline=(char.base)]{
            \node[shape=circle,draw,inner sep=0.5pt] (char) {#1};}}

\newcommand{\framework}{\textit{NsPC}\xspace}

\newcommand{\llms}{\textit{LLMs}\xspace}

\newcommand{\lcm}{LCM\xspace}
\newcommand{\lcms}{LCMs\xspace}

%% file: sections/0_abstract.tex
\begin{abstract}
Recent advancements in Large Language Models (\llms) have paved the way for Large Code Models (\lcms), enabling automation in complex software engineering tasks, such as code generation, software testing, and program comprehension, among others. Tools like GitHub Copilot and ChatGPT have shown substantial benefits in supporting developers across various practices. However, the ambition to scale these models to trillion-parameter sizes, exemplified by GPT-4, poses significant challenges that limit the usage of Artificial Intelligence (AI)-based systems powered by large Deep Learning (DL) models. These include rising computational demands for training and deployment and issues related to trustworthiness, bias, and interpretability. Such factors can make managing these models impractical for many organizations, while their ``black-box'' nature undermines key aspects, including transparency and accountability. In this paper, we question the prevailing assumption that increasing model parameters is always the optimal path forward, provided there is sufficient new data to learn additional patterns. In particular, we advocate 
for a \textbf{N}eurosymbolic research direction that combines the strengths of existing DL techniques (\eg LLMs) with traditional symbolic methods--renowned for their reliability, speed, and determinism. To this end, we outline the core features and present preliminary results for our envisioned approach, aimed at establishing the first \textbf{N}eurosymbolic \textbf{P}rogram \textbf{C}omprehension (\framework) framework to aid in identifying defective code components.

\end{abstract}


%% file: sections/1_introduction.tex
\section{Introduction}
\label{sec:introduction}
There is no doubt that the recent rise of Large Code Models (\lcms) has revolutionized the automation of Software Engineering (SE) activities. 
To understand \emph{why}, \emph{when}, and \emph{how} this transformation occurred, we must narrow down our analysis to two key aspects that have contributed significantly to this revolution: (i) the availability of large, text-rich datasets, which provide the foundational knowledge required for training these models, and (ii) the increasing scale of deep learning (DL) architectures, with models now boasting trillions of parameters (\eg GPT-4 \cite{openai2024gpt4}). These two elements together have not only expanded the ability of models to embed and generalize vast amounts of programming knowledge but also facilitated their capacity to capture peculiar elements within the code, including intricate patterns, structures, and relationships. This duality (\ie large corpus and models) laid down the groundwork for achieving new levels of automation in SE, that were once thought to be beyond reach.

In this regard, tools, functioning as ``artificial collaborators'', such as GitHub Copilot \cite{copilot} and ChatGPT \cite{chatgpt}, have been effective in assisting and supporting developers in multiple phases of the software development lifecycle \cite{white2023chatgpt,nashid2023retrieval,watanabe2024use} as well as enhancing their understanding of code \cite{khojah2024beyond,dakhel2023github}.

While recent literature has presented the various and multifaceted possibilities of AI methods for software engineering activities \cite{hou2023large}, the ``no free lunch'' theorem reminds us that these benefits come at a cost. In particular, as models continue to grow in complexity and scale, the computational demands for training and maintaining them have become a significant burden \cite{stojkovic2024greener}. Also, concerns about bias, trustworthiness, and interpretability in large DL models such as \lcms, highlight a significant roadblock, preventing further advancement.



In this paper, we challenge the prevailing belief that scaling up models indefinitely is the path forward for every domain where AI-driven methods are deployed, including SE. 
To this end, Villalobos \etal \cite{villalobosposition} recently challenged the assumption that Large Language Models (\llms) can continue to learn effectively from existing data. They noted that society is approaching a point where the amount of relevant information available for \llms to learn from will be nearly exhausted, an event projected to occur between 2026 and 2032. In other words, we are nearing a critical threshold where the size of these models--counted in terms of parameters--could outstrip the volume of meaningful data available for processing.

Given this state of affairs, we ask: \emph{``What if we take a step back now to move two steps forward later?''} In other words, we have hit the limits of improvement through sheer model scaling, making it necessary to reconsider the dominant paradigm that has fueled innovations in the past decade.

With this in mind, our overarching goal is to \textbf{\emph{develop a new framework that harnesses the probabilistic capabilities of \llms while seamlessly integrating traditional symbolic rules}}. This combination enhances interpretability, ensures deterministic reasoning, and overcomes the inherent limitations of purely probabilistic approaches--that as seen--are increasingly plateauing.

As a first step towards this endeavor, we focus on \textit{vulnerability detection}, a critical task in software security that heavily depends on program comprehension. Understanding how code is structured and behaves is essential for identifying security weaknesses, as detecting vulnerabilities requires the ability to analyze and reason about code effectively. This understanding also plays a key role in related tasks such as debugging, refactoring, and secure software maintenance. To support these efforts, we propose the \textbf{N}eurosymbolic \textbf{P}rogram \textbf{C}omprehension (\framework) paradigm, which combines \lcms with symbolic reasoning to equip developers with more powerful tools for identifying and addressing insecure code.




In this research, we present preliminary analyses and results aimed at developing the first \framework approach for vulnerability detection \cite{zhou2019devign}, leveraging SHAP \cite{NIPS2017_Lundberg}, an interpretability method that generates local explanations for individual predictions (\eg determining whether a code component is affected by a vulnerability). By using SHAP values, we envision to uncover underlying patterns, translate them into symbolic rules, and seamlessly integrate these rules into the \lcm. To the best of our knowledge, this is the first documented attempt to embed a symbolic layer into the probabilistic framework of \lcms for program comprehension, introducing a promising direction for automating software engineering practices. This novel approach prioritizes not only peak performance but also interpretability and transparency, paving the way for future methods in the SE domain.

%% file: sections/2_background.tex
\section{Background}
\label{sec:background}

Shapley Additive exPlanations (SHAP) \cite{NIPS2017_Lundberg} is a technique for estimating each feature's contribution to the output $y$ of a deep learning model $f(x)$. Rooted in cooperative game theory, SHAP is based on Shapley values, introduced by Lloyd Shapley as a method for fairly distributing payouts among participants in cooperative games \cite{shapley:book1952}. SHAP values correspond to the Shapley values of a conditional expectation function derived from the model, capturing feature interactions and dependencies to provide robust explanations.

In practice, SHAP isolates the impact of individual features ($w_i \in x$) on the model's output while accounting for the influence of other features ($x \setminus w_i$). It calculates the average difference in predictions when a feature is included versus excluded, offering insights into how features influence the model’s decisions. SHAP is applicable across various models, including tree-based \cite{kumar_shapley_problems_2020} and neural network models \cite{ahn_shapley_www_2024}, enabling researchers to identify key predictors and analyze model behavior. Its flexibility and strong theoretical foundation make SHAP invaluable for post-hoc interpretability \cite{sundararajan_many_shapley_2020}, particularly in applications requiring both accuracy and interpretability, such as medical diagnostics \cite{stiglic_health_interpretability_2020}, financial risk assessment \cite{barnes_finance_interpret} and software engineering tasks, as explored in this study.

%% file: sections/2_approach.tex
\section{Methodology}
\label{sec:approach}

In this section, we present the \textbf{N}eurosymbolic \textbf{P}rogram \textbf{C}omprehension (\framework) framework, which leverages SHAP values (refer to \secref{sec:background}) to interpret and guide model predictions. We first describe our approach to identifying patterns in SHAP values for input features. Next, we explain how these patterns are transformed into symbolic rules to improve model performance, particularly in scenarios with low prediction confidence.

\begin{figure}[ht]
		\centering
  \vspace{-1.5em}
  \includegraphics[width=0.45\textwidth]{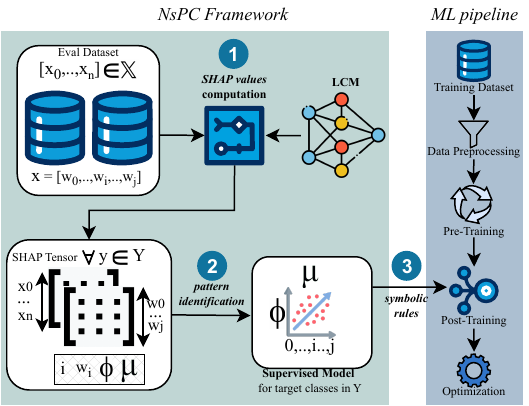}
		\caption{Description of \framework framework as a sequence of steps.}
    \label{fig:npc_pipeline}
    \vspace{-1em}
\end{figure}

\input{tables/results}

\subsection{Pattern Identification}
\label{sec:npc_pattern_identification}

Drawing inspiration from probing classifier techniques widely used in NLP \cite{hewitt_designing_2019} and SE \cite{lopez_ast-probe_2022, troshin_probing_2022}, our framework leverages supervised machine learning techniques to identify patterns in the SHAP values computed for specific predictions in classification tasks. Probing techniques work by examining the latent representations of a model to determine the extent to which specific types of information are encoded. Specifically, a supervised model (\eg classifier) is trained to predict properties of interest from the neural network's hidden representations \cite{belinkov_probing_2021}. In the context of our framework, we propose training classifiers to predict target classes from SHAP value distributions enabling the formulation of symbolic rules, as illustrated in \figref{fig:npc_pipeline}.

First, given a set of inputs $\mathbb{X}$ that the \lcm predicts as belonging to a specific class $y \in Y$ (\eg Secure/Insecure), we compute SHAP values ($\phi$) for each input $x \in \mathbb{X}$. The SHAP values are calculated relative to the expected predicted class: $y = \mathbb{E}[f(\mathbb{X})]$. Inspired by syntax decomposition \cite{syntax_capabilities, palacio_towards_2024, docode}, we apply an alignment function $\delta(w_i): w_i \to \mu \in \mathbb{M}$ to tag tokens $w_i \in x$ with meaningful AST types $\mathbb{M}$, defined by the programming language grammar. This process produces a SHAP tensor for each target class: ${(i, w_i, \phi_i, \mu_i)}$, where $i$ is the position, $w_i$ is the token, $\phi_i$ is the SHAP value, and $\mu_i$ is the associated AST type. The entire process is depicted in region \circled{1} of \figref{fig:npc_pipeline}.

After computing the SHAP tensors for each target class in $Y$, we merge them and group the $\phi$ values by the AST tag associated with their corresponding tokens. We define position ranges as $[a, b], \quad 0 \leq a \leq b \leq \max{|x|: x \in \mathbb{X}}$. For each range, we train a supervised model (\eg logistic regression, decision tree, random forest) to identify curves that best capture the relationship between $\phi$ values and feature positions. Curves with an accuracy exceeding $60\%$ and a well-defined decision boundary for the target class (\ie intersection with the x-axis) provide evidence of patterns in specific AST type positions where SHAP values influence the model's decisions. The computed curves allow us to identify regions and position ranges where a feature’s $\phi$ value (\ie SHAP value corresponding to a specific AST node) consistently influences the overall prediction of the expected outputs either positively or negatively.

\subsection{Symbolic Rules}
From the identified patterns in SHAP value distributions, we derive symbolic rules encapsulating feature structures that align with expected model predictions. These rules consist of two parts: (i) configurations positively correlated with the predicted label, forming symbolic rules for correctly predicted patterns, and; (ii) complementary rules for configurations linked to lower prediction reliability, enabling targeted model adjustments in uncertain cases. We derive these rules by grouping SHAP-influential features within each type $\mu \in \mathbb{M}$ and formulating conditions based on both feature presence and SHAP value contributions. For instance, if a feature linked to an AST node consistently shows high SHAP values for insecure code at the input's start, it may represent a necessary condition for an \textbf{\textit{insecure}} prediction in the rule. As illustrated in region \circled{3} of \figref{fig:npc_pipeline}, the derived symbolic rules can be applied during the post-training stage of an ML pipeline, for instance, in supervised fine-tuning and knowledge distillation to facilitate knowledge transfer between models.

%% file: tables/results.tex
\begin{table*}[]

\centering
\caption{Logistic Regression Results by Type and Position Range. Cells with a gray background indicate the position ranges where the logistic regression model suggests the presence of a rule.}
\label{tab:results}
\vspace{-0.5em}
\scalebox{0.85}{

\setlength{\tabcolsep}{4pt} 

\begin{tabular}{lllllllllllllllllll}
 &
   &
  \multicolumn{2}{c}{{[}0-50{]}} &
   &
  \multicolumn{2}{c}{{[}51-100{]}} &
   &
  \multicolumn{2}{c}{{[}101-150{]}} &
   &
  \multicolumn{2}{c}{{[}151-200{]}} &
   &
  \multicolumn{2}{c}{{[}201-250{]}} &
   &
  \multicolumn{2}{c}{{[}251-300{]}} \\
\textit{\textbf{AST Type}} &
   &
  \textit{\textbf{accuracy}} &
  \textit{\textbf{x-int}} &
   &
  \textit{\textbf{accuracy}} &
  \textit{\textbf{x-int}} &
   &
  \textit{\textbf{accuracy}} &
  \textit{\textbf{x-int}} &
   &
  \textit{\textbf{accuracy}} &
  \textit{\textbf{x-int}} &
   &
  \textit{\textbf{accuracy}} &
  \textit{\textbf{x-int}} &
   &
  \textit{\textbf{accuracy}} &
  \textbf{x-int} \\ \hline
\textit{identifier} &
   &
  0.55 &
  24 &
   &
  0.52 &
  - &
   &
  0.51 &
  - &
   &
  0.51 &
  - &
   &
  0.51 &
  213 &
   &
  0.54 &
  - \\
\textit{type} &
   &
  0.49 &
  19 &
   &
  0.49 &
  - &
   &
  0.44 &
  135 &
   &
  0.62 &
  - &
   &
  0.33 &
  217 &
   &
  \cellcolor[HTML]{C0C0C0}0.67 &
  \cellcolor[HTML]{C0C0C0}299 \\
\textit{punctuation} &
   &
  \cellcolor[HTML]{C0C0C0}0.61 &
  \cellcolor[HTML]{C0C0C0}45 &
   &
  0.60 &
  - &
   &
  0.62 &
  - &
   &
  0.58 &
  - &
   &
  0.55 &
  - &
   &
  0.51 &
  - \\
\textit{access\_modifiers} &
   &
  0.66 &
  - &
   &
  - &
  - &
   &
  - &
  - &
   &
  - &
  - &
   &
  - &
  - &
   &
  - &
  - \\
\textit{operator} &
   &
  \cellcolor[HTML]{C0C0C0}0.74 &
  \cellcolor[HTML]{C0C0C0}38 &
   &
  0.51 &
  54 &
   &
  0.54 &
  - &
   &
  0.51 &
  - &
   &
  0.55 &
  - &
   &
  \cellcolor[HTML]{C0C0C0}0.59 &
  \cellcolor[HTML]{C0C0C0}280 \\
\textit{literal} &
   &
  \cellcolor[HTML]{C0C0C0}0.60 &
  \cellcolor[HTML]{C0C0C0}43 &
   &
  0.55 &
  - &
   &
  0.53 &
  - &
   &
  0.58 &
  - &
   &
  0.53 &
  - &
   &
  0.48 &
  - \\
\textit{primitive} &
   &
  0.50 &
  20 &
   &
  \cellcolor[HTML]{C0C0C0}0.57 &
  \cellcolor[HTML]{C0C0C0}77 &
   &
  0.33 &
  139 &
   &
  - &
  - &
   &
  - &
  - &
   &
  - &
  - \\
\textit{comment} &
   &
  0.47 &
  11 &
   &
  0.61 &
  - &
   &
  0.52 &
  - &
   &
  0.57 &
  - &
   &
  0.60 &
  - &
   &
  0.69 &
  - \\ \hline
\end{tabular}

} 
\vspace{-1em}
\end{table*}

%% file: sections/3_case_study.tex
\section{Case Study}
\label{sec:case_study}

To demonstrate the practical application of NsPC, we conducted a case study to identify SHAP value patterns in the context of insecure code detection (\ie binary classification task) using Java code snippets. This study aimed to address the following research question:

\begin{enumerate}[label=\textbf{RQ$_{\arabic*}$}, ref=\textbf{RQ$_{\arabic*}$}, wide, labelindent=5pt]\setlength{\itemsep}{0.2em}
      \item \label{rq:neuro_symbolic_rules} {\textbf{[Symbolic rules from SHAP]} To what extent SHAP values enable the definition of symbolic rules?}
\end{enumerate}

\textbf{Selected \lcm.} For our analysis, we selected CodeBERT \cite{feng_codebert_2020}, fine-tuned for detecting insecure code snippets \footnote{https://huggingface.co/mrm8488/codebert-base-finetuned-detect-insecure-code} as BERT-like architectures are widely adopted in SE for classification tasks \cite{10.1145/3528588.3528660,9492202,ardimento2020using,chochlov2022using}. Specifically, we focus on a binary classification task, where the presence of insecure code in a code snippet is treated as the \textit{"positive"} class prediction, while the absence of insecure code is the \textit{negative}. The selected model, trained on the Devign \cite{zhou_devign_2019} (CodeXGLUE--Defect Detection \cite{lu_codexglue_2021}), features a vocabulary size of $50,265$ and comprises $12$ hidden layers with attention heads. The model was deployed on an Ubuntu 20.04 system with an AMD EPYC 7532 32-Core CPU, an NVIDIA A100 GPU with 40GB VRAM, and 1TB of RAM.

\textbf{Evaluation Dataset.} For evaluation, we used the validation split of the CodeXGLUE dataset for Defect Detection. Specifically, we created two smaller datasets by splitting the datapoints based on the target class (\ie positive and negative). To align with the token limit defined by the selected model, we restricted each data point to a maximum of $500$ tokens. The resulting datasets included a total of $300$ confirmed insecure datapoints for the positive target class and $300$ datapoints free of insecure code for the negative target class.

\textbf{Evaluation Methodology}. To address \ref{rq:neuro_symbolic_rules}, we applied \framework to compute SHAP tensors for each of the two evaluation datasets (refer to \secref{sec:npc_pattern_identification}). We analyzed these tensors by defining six position ranges, considering a maximum token length of $300$ per snippet. Additionally, we trained logistic regression models to compute decision boundaries within these ranges for the two possible classes: \textbf{\textit{secure}} and \textbf{\textit{insecure}}.

\subsection{Results \& Discussion}
\label{sec:case_study_results_discussion}

\begin{figure*}[ht]
		\centering
  \includegraphics[width=1\textwidth]{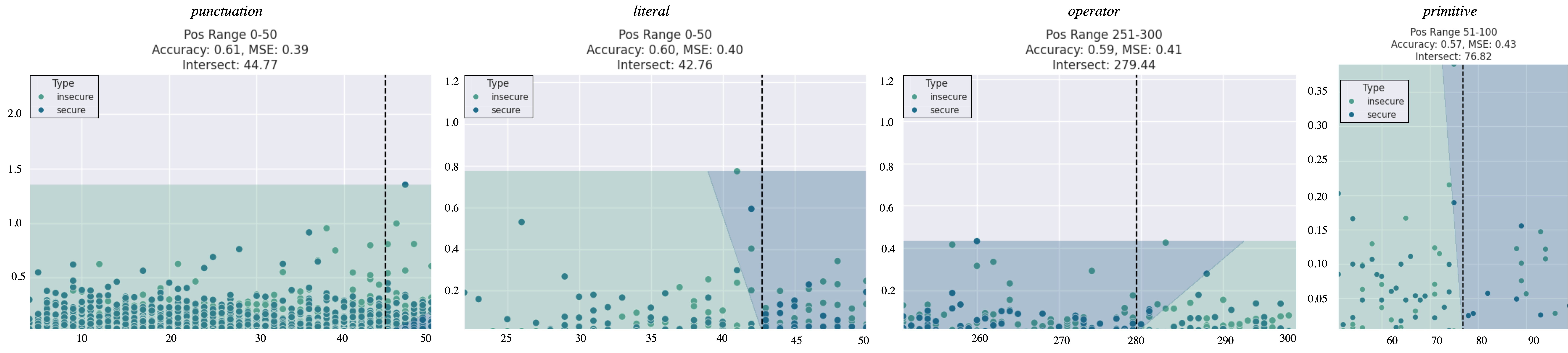}
		\caption{Examples of logistic regression models suggesting the presence of a pattern in each position range per type of AST element.}
    \label{fig:rules_examples}
    \vspace{-1em}
\end{figure*}

\tabref{tab:results} summarizes the results of the trained logistic regression models for each position range and identified AST type. The trained logistic regression model surpassed the $60\%$ accuracy threshold and exhibited a clear decision boundary for the two possible outcomes only for the AST types \textit{punctuation}, \textit{operator}, \textit{literal}, \textit{type}, and \textit{primitive}. For instance, as illustrated in \figref{fig:rules_examples}, if a snippet contains a \textit{literal} token within positions $[0-43]$, there is a high probability that the snippet will be classified as \textbf{\textit{insecure}}. Similarly, if a snippet contains an \textit{operator} within positions $[251-280]$, there is a high probability that the snippet will be classified as \textbf{\textit{secure}}.


These patterns reflect meaningful correlations arising from the underlying dataset and programming conventions. For instance, literal tokens often appear early in code snippets due to the prevalence of hardcoded values, initialization blocks, or function arguments, which are common in insecure patterns. Conversely, \textit{operator} tokens in later positions typically belong to logical constructs or functional operations, often associated with structured and secure code. 
We capitalize on these patterns to present compelling evidence supporting the instantiation of the \emph{NsPC} framework to identify symbolic rules for the selected \lcm (\ie fine-tuned CodeBERT).



However, as the pattern identification relies on SHAP values computed specifically for this model, the evidence obtained is not sufficient to generalize these rules to other models, tasks, or datasets, highlighting the need for further research. Nevertheless, this study represents a foundational step toward introducing the first \framework in the literature aimed at supporting program comprehension tasks, with a particular focus on vulnerability detection.


\begin{boxK}
    \textit{\ref{rq:neuro_symbolic_rules}} \textbf{[Neurosymbolic Component]}: Using the proposed \framework framework, we identified meaningful insecure-prone patterns within specific position ranges, which facilitated the definition of symbolic rules for detecting \textbf{secure} and \textbf{insecure} code snippets. The patterns reveal that tokens from certain AST types in particular positions have a significant impact on the model’s predictions.
\end{boxK}

%% file: sections/4_related.tex
\section{Related Work}
\label{sec:related_work}


In this section, we present an overview of studies relevant to this paper, including (i) the interpretability of models for SE and; (ii) applications of neurosymbolic AI for SE.

\textbf{Interpretability of Models for SE:} Chen \etal \cite{chen_cat-probing_2022} introduce CAT-probing, a method to quantitatively interpret how pre-trained models (CodePTMs) for programming languages capture the structural properties of code. They highlight that the middle layers in models may significantly influence transfer of general structural knowledge, while later layers refine task-specific knowledge. Anand \etal \cite{anand_critical_2024} approach interpretability of code \llms (cLLMs) via attention analysis and show that attention maps of cLLMs fail to encode syntactic-identifier relations. Bui \etal \cite{bui_autofocus_2019} aim to enhance the interpretability of attention-based models for code by adapting code perturbations to evaluate the meaningful code elements. Other research works proposed interpretability techniques by applying the principles of information storage \cite{haider_looking_2024},  AST-probe \cite{majdinasab_deepcodeprobe_2024}, and syntactic structures combined with prediction confidence \cite{palacio_towards_2024}.

\textbf{Neurosymbolic AI in SE:} Princis \etal \cite{princis_sql_2024} integrate symbolic reasoning techniques into \llms to improve SQL query generation. This hybrid system leverages symbolic checks for query validation and repair during the generation process. To achieve this, the system employs partial query evaluation and early elimination of invalid queries, significantly improving runtime and accuracy. The study does not explore the interpretability of this hybrid system. 

Arakelyan \etal \cite{arakelyan_ns3_2022} combine neural and symbolic methods to improve the multi-step reasoning and compositional querying abilities of semantic code search (SCS) systems. The approach utilizes rule-based parsing of the natural language queries to identify matches between the parsed query components and code snippets. The rules, however, are manually created by the authors and might not generalize well for other natural and programming languages.

Jana \etal \cite{jana_cotran_2024} present CoTran, an LLM-based neurosymbolic system for translating code between programming languages. The proposed system leverages a \textit{symbolic execution feedback} to ensure functional equivalence of translated code. The code translation is available between Java and Python languages. Integration of the symbolic component improves the system's ability to maintain the original code's logic and leads to more robust and reliable translations.

There are also works on the applications of neurosymbolic AI techniques in program synthesis \cite{parisotto_neuro-symbolic_2016} \cite{bosnjak_programming_2017}, representation learning \cite{allamanis_learning_2017}, error correction \cite{xue_interpretable_2024}, semantic code repair \cite{devlin_semantic_2017}, and bug fixing \cite{hu_fix_2022}.




%% file: sections/5_future_plan.tex
\section{Future Plans}
\label{sec:future_plan}

In this paper, we presented our framework designed to enhance the capabilities of \lcms through the definition of a deterministic layer built upon symbolic rules. By leveraging interpretability techniques such as SHAP, our approach identifies patterns in model predictions, which can be formalized into symbolic rules. We believe that interpretability techniques not only provide valuable insights into model behavior but also serve as a foundation for defining rules that improve both the transparency and performance of \lcms, particularly in tasks requiring high reliability and explainability. In other words, we are challenging the canonical paradigm that has dominated software engineering automation over the past decade, where the predictive capabilities of machine learning methods, particularly deep neural networks, have streamlined various SE-related practices


As next steps, we aim to address two fundamental key areas to further refine and expand our framework. First, we seek to establish a more rigorous mathematical foundation for our framework to formalize its theoretical underpinnings and improve its reliability and generalizability in diverse applications that extend beyond classification tasks. 
Second,  we aim to incorporate human validation of the derived symbolic rules to ensure their correctness, interpretability, and practical relevance, thereby bridging the gap between automated rule generation and real-world applicability.

